\documentclass[prb,superscriptaddress,showpacs]{revtex4}
\usepackage{epsfig}
\begin{document}
\title{Double quantum dot turnstile as an electron spin entangler}
\author{Xuedong Hu}
\affiliation{Department of Physics, University at Buffalo, SUNY,
Buffalo, NY 14260-1500}
\affiliation{Condensed Matter Theory Center, Department of Physics, 
University of Maryland, College Park, MD 20742-4111}
\author{S. Das Sarma}
\affiliation{Condensed Matter Theory Center, Department of Physics,
University of Maryland, College Park, MD 20742-4111}
\date{\today}
\begin{abstract}
We study the conditions for a double quantum dot system to work as a
reliable electron spin entangler, and the efficiency of a beam splitter as a
detector for the resulting entangled electron pairs.  In particular, we focus
on the relative strengths of the tunneling matrix elements, the applied bias
and gate voltage, the necessity of time-dependent input/output barriers, and
the consequence of considering wavepacket states for the electrons as they
leave the double dot to enter the beam splitter.  We show that a double
quantum dot turnstile is, in principle, an efficient electron spin entangler
or entanglement filter because of the exchange coupling between the dots and
the tunable input/output potential barriers, provided certain conditions are
satisfied in the experimental set-up.
\end{abstract}
\maketitle

\section{Introduction}
\label{sect-intro}

Experimental and theoretical studies in quantum computing and quantum
information processing have shown that there exist natural resources in
the quantum regime\cite{Chuang} such as quantum superposition and
entanglement that can be exploited to provide additional computing power.  In
particular, the study of quantum entanglement has attracted wide spread
attention because of its direct relevance to quantum computation and its
implications to the foundations of quantum mechanics.\cite{Chuang,review}  
Many physical systems ranging from atomic and optical to solid state have been
proposed as potential candidates for quantum information processing and for
providing insights to their inherent quantum mechanical properties. 
Specifically, localized spins (electron or nuclear) trapped in solid state
host materials (particularly semiconductors) have been considered as good
candidates for these purposes because of their relatively isolated (from
their environment) nature.\cite{LD,Vrijen,Kane1,Privman,Skinner}  In this
context, the creation and the detection of electron spin entanglement become
critically important tasks, and are the subject matter of interest to us in
this paper.

Quantum entanglement is a manifestly non-classical property of the quantum
state of a composite system (e.g. two or more particles) where the entangled
composite state cannot be decomposed into a product of the individual states
of local constituents, and as such the constituents are ``entangled" no matter
how far they are separated spatially.  The classic example is the spin
singlet ``EPR" state of two spin-1/2 fermions, where, no matter how far apart
the two particles are spatially, a measurement of the spin of one
particle completely determines the quantum spin state of the
other (provided, of course, the spin singlet state of the two-particle state
is preserved coherently up to the measurement process, i.e. no decoherence
takes place until measurement).  

Entanglement leads to specific nonlocal (and nontrivially nonclassical)
correlations in the measured properties of the individual constituents, which
are typically expressed in terms of the celebrated Bell's inequalities.  A
violation of Bell's inequalities in correlation measurements of the
constituents indicates the presence of entanglement in the technical sense
(i.e. shows that the state of one constituent is inextricably and nonlocally
quantum-mechanically entangled with the state of the other constituent no
matter how far spatially apart they may be).  Direct demonstration of
entanglement (as manifested in the violation of Bell's inequalities) have so
far been limited mostly to experiments involving
photons\cite{Aspect,Tittel,Weihs}
because entangled photons are easy to produce in laboratories using the
parametric down conversion processes in optical nonlinear crystals, and
photons have very long coherence lengths since they are extremely weakly
interacting objects.  However, from a classical perspective at least (i.e.
when considering light as an all pervading wave rather than a collection of
quantized photons), nonlocal entanglement manifestation of light waves is not
a particularly shocking situation as it would be with classically massive
objects such as atoms\cite{Rowe} and electrons, which are classically purely
particle-like, making any nonlocal classical correlation impossible.

Entanglement (in the technical Bell's inequalities or EPR sense) has never
been experimentally demonstrated in any condensed matter systems.  
Because of strong interactions inherently present in all solid state
systems, the ground state is often an interacting many-body state (e.g. the
strongly correlated Laughlin state for the fractional quantum Hall system,
the Bethe ansatz singlet ground state for one-dimensional spin chains, the
BCS superconducting ground state) where the collective state is highly
'entangled' in the sense that it cannot be written as a mean field product
state of one electron Hartree wavefunctions.\cite{HF_comment,Schliemann} 
While such strongly coupled states (the Laughlin state in the fractional
quantum Hall system being a classic example) are intrinsically entangled by
definition, in general they are unsuitable for the observation of
entanglement in the technical sense of EPR/Bell's inequalities, since turning
off the interactions and spatially separating such entangled electrons so as
to carry out correlation measurements
are essentially impossible in the strongly interacting solid state
environment.  We mention, however, that perhaps more thinking and research
should go into exploring the possibility of using a strongly coupled quantum
many-body ground state such as the Laughlin state to experimentally
demonstrate entanglement correlations.  

Experimental demonstration of solid state entanglement should entail the
following minimal steps as a matter of principle: (1) Two (or more) particles
(or sub-systems) interact to form an entangled pair; (2) this entangled pair
of constituents is then physically separated from the specific solid state
environment;\cite{weak_int} (3) the two constituents are spatially separated 
and their interaction {\it turned off} while the entangled two-particle state
is preserved;\cite{weak_int,spatial_sep} (4) when the two constituents are
sufficiently spatially separated, some suitable properties of each are
measured in a correlated manner; (5) a study of correlation between these
local measurements on each constituent, 
provided they violate Bell type inequalities, establishes entanglement between
the constituents.  We emphasize that the hardest steps in the experimental
observation of solid state entanglement are the steps two and three above 
because it is extremely hard to spatially separate electrons in solids in a
controlled manner without causing decoherence.  The problem in
observing solid state entanglement is really too much (and not too little)
entanglement---electrons in solids are intrinsically coupled or entangled
with each other (and with the environment) and extracting a pair of
constituents and spatially separating them in order to observe entangled
correlations within the pair is very difficult, if not impossible.

A possibly suitable method to observe solid state electronic entanglement is
to use transport or tunneling techniques, which naturally allow flow or
movement of particles enabling spatial separation, the essential ingredient
for observing entangled correlations.  For example, one could extract a
Cooper pair from a superconductor, which is in an entangled (delocalized)
singlet state in general.  Then the two electrons in this Cooper pair have to
be spatially separated somehow (using coherent tunneling or transport through
various leads) in a ballistic manner (i.e. without any decoherence
whatsoever) so that correlation measurements on each electron of the pair can
be carried out locally on each system without any interaction with its
distant partner. 
Clearly, decoherence will be a severe problem in such experiments as each
electron of the singlet pair will necessarily interact strongly with the
environment (i.e. all ``other" electrons in the system and in the leads,
phonons, magnetic and other impurities).  In spite of all these problems,
many approaches for creating/detecting entanglement in solid state systems 
(involving, for example, delocalized Cooper
pairs\cite{DFHZ,Lesovik,Cht,Recher1,Recher2,Bena}, localized quantum dot or
quantum wire electron singlets\cite{Ionicioiu,WOliver,Saraga},
beam splitters\cite{BLS,Costa,Bose,Samuelsson,Been1}, etc) have been
theoretically proposed in
the literature to study spin entanglement properties in a solid state
environment, and in this paper we study in some depth a proposal in which
electron spin entanglement is generated in a coupled double quantum dot and
detected through a beam splitter, where measurements on the scattered current
noise and correlation have been shown to exhibit possible signatures
of an entangled electron spin singlet state in semiconductors.\cite{BLS} 
This detection proposal is still far from a Bell-type measurement, but it is a
necessary first step as it deals with correlations between electrons that
have been extracted from their entangler and are already spatially separated.  

In this paper, we first discuss in Section~\ref{sect-DQD} and
\ref{sect-problem} how a double quantum dot might be used to generate pairs
of entangled electrons and what problems may affect the performance of a
fixed barrier double dot system as a spin entangler.  Then in
Section~\ref{sect-turnstile} we introduce the dynamic turnstile configuration
in order to create a regulated source
of entangled electrons, and in Section~\ref{sect-wavepacket} we discuss how
the wavepacket nature of the electrons reduces the signature of electron
entanglement in a beam splitter.  One key aspect of our work is to analyze
the proposed noise/correlation experiment of Ref.~\onlinecite{BLS} in terms
of particle-like wavepacket electronic states.  The other key aspect is to
show that it will be essentially impossible to satisfy all the stringent
temporal restrictions (e.g. tunneling rates for the electrons to go in and
out of the dots, the decoherence rate, etc) needed for the success of the
proposed detection scheme using a static barrier double quantum dot
system---we therefore suggest and analyze a situation where the double dot
input/output barriers will operate in a dynamical time-dependent turnstile
mode, first allowing entanglement to occur in the double dot through the
exchange interaction and then enabling the extraction of the entangled
electrons and their detection through noise cross-correlation measurements in
the output leads.

\section{Double dot as an entanglement filter: Fixed Barriers}
\label{sect-DQD}

The basic idea underlying the use of a coupled double quantum dot system as an
electron spin entangler has already been introduced and
discussed.\cite{LD,BLD,HD1,HD2}  It is now
well-established that an exchange-coupled double quantum dot could act as an
artificial molecule where individual electrons on one dot are ``entangled"
with electrons on the other in molecular singlet (or triplet) states,
similar to the corresponding situation in real molecules.  For example, a
two-electron double dot with one electron on each dot (and a suitable gate to
control the exchange coupling between the two dots by electrically tuning the
overlap of the electron wavefunctions in the two dots) is the artificial
analog of the H$_2$ molecules on a 10 to 100 nm size and 0.1 to 1 meV
energy scale (although the details are qualitatively different since the
``atomic" electronic confinement in the quantum dot case is the approximately
parabolic external confinement potential imposed by electrostatic gates
in contrast to the Coulombic nuclear confinement potential in real atoms and
molecules).  In a two-electron double quantum dot the ground state (with one
electron in each dot), in the absence of any external magnetic field, is the
exchange-coupled and manifestly entangled spin singlet state.\cite{BLD,HD1} 
In multielectron (i.e. each dot containing more than one electron) double
quantum dots\cite{HD2} the ground state, even in the absence of any applied
magnetic field, could be either a singlet or a triplet or other more
complicated states depending on the details of the system (e.g. the number of
electrons, the confinement potential, the exchange gate behavior) just as in
real many-electron molecules where the ground state may be a spin triplet
state.  

\begin{figure}
\centerline{
\epsfxsize=3.0in
\epsfbox{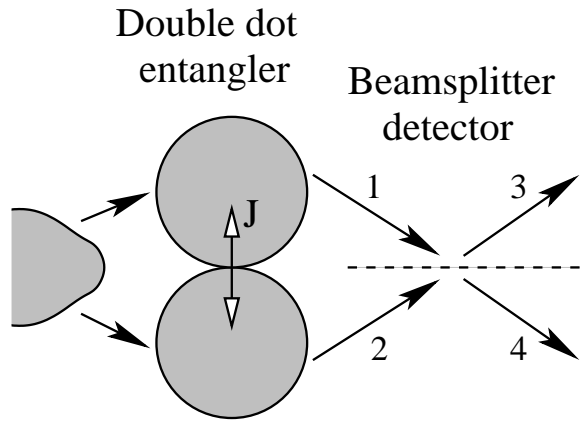}}
\vspace*{0.1in}
\protect\caption[]
{\sloppy{
Schematic diagram of a double quantum dot as an electron-spin-entanglement
filter.  An electron beam splitter \cite{Liu} is included as a potential
detector for the spin entanglement \cite{BLS}.
}}
\label{fig-Sch}
\end{figure}

The idea is therefore to introduce electrons into the double dot structure
through well-controlled input barrier in order for them to populate the spin
singlet states and then to controllably extract the entangled electron pairs
from the double dot system through output barriers into transport leads.  Then
the entanglement has to be detected in a suitable transport measurement
which, as already proposed in Ref.~\onlinecite{BLS}, could be a current noise
and/or cross-correlation measurement between the electrons in the entangled
pair through a beam splitter (Fig.~\ref{fig-Sch}).  The basic detection idea
is simple conceptually but extremely difficult to implement experimentally. 
We note that the proposed current noise measurement satisfies the key
criterion for establishing entanglement in the sense that these measurements
involve local measurements for each constituent which are spatially separated
from each other (and are no longer interacting with each other through the
direct exchange coupling).  Entanglement manifests itself in the correlations
between local observations on spatially separated constituents.  By contrast,
a direct spectroscopic observation of the singlet state in the double quantum
dot system itself, for example by observing the singlet-triplet energy level
splitting (or the symmetric-antisymmetric gap), is in our view not direct
evidence for entanglement, but evidence for the presence of inter-dot
electronic coupling or wavefunction overlap.  It is only by separating the
two electrons, as shown schematically in Fig.~\ref{fig-Sch}, one could hope
to demonstrate entanglement.  (Demonstration of, or evidence for, entanglement
necessarily requires that the local interaction originally creating the
entangled state has been turned off.)

We consider a horizontally coupled double quantum dot (QD) configuration as
shown in Fig.~\ref{fig-Sch}, where the two dots are connected to the same
source electrode but different drain electrodes, so that the pair of
electrons can be separated after passing through the double dot.  As is
evident from the schematics, such a coupled double dot has several knobs we
can tune to adjust its properties because modern lithographic technologies
allow various gates and leads to be put on the double dot structure in a
controllable manner.  These include the bias voltage applied
between the source and drain electrodes, the gate voltages applied to the two
quantum dots, the inter-dot tunnel coupling, and the dot-electrode
couplings.  Parallelly accessible double quantum dots have already been
studied experimentally,\cite{parallel} so that the configuration discussed
here should be feasible for future experimental exploration.  Below we will
first analyze some requirements imposed by our objective of using the double
dot as an entangler for electron spins.

We assume {\it a priori} that the tunnel coupling between the double dot is
stronger than the tunnel coupling between the double dot and the source and
drain electrodes, so that electrons inside the double dot can form a
molecular state.  This is an absolutely necessary condition for our proposed
spin entangler to work.  This assumption also dictates that only one source
electrode is needed because the electrons in the reservoir(s) would see the
double dot as a single (strongly tunnel-coupled) entity.  For the case of
weak inter-dot tunnel coupling, we can consider the limit of vanishing
coupling to understand why such a situation is not conducive to our objective
of creating spin-entangled electron pairs.  If there is no tunneling or
wavefunction overlap between the two dots, the only possible inter-dot
interaction is the classical Coulomb long range coupling.  The two streams
of electrons passing through the two quantum dots would have no spin
correlations in this situation.  The currents will be correlated because of
the Coulomb blockade: when one of the QDs is occupied, the other would tend
to stay empty to minimize the total energy of the system (depending on the
applied gate voltage).  The situation is quite similar to what has been
studied in Ref.~\onlinecite{Hoffman}, and is not useful for the purpose of
entangling electron spins.  (Here is in some sense a rather simple example of
a coupled system ``without entanglement", precisely speaking ``without spin
entanglement".)

For simplicity, we consider a situation where the double dot is empty 
initially.  This condition can be achieved by tuning the gate voltages to
empty the two QDs, which has been achieved experimentally in both vertical 
and horizontal QDs.\cite{vertical,horizontal,Ciorga2,Elzerman}  If the QDs are
not initially empty, exchange coupling can still be established.\cite{HD2} 
However, chaotic behavior sets in as the number of electrons increases, so
that the exchange coupling would be very sensitive to the energy levels
involved and can vary uncontrollably in a wide range.  To avoid this
sensitivity to the control gate voltages, we will limit ourselves to the
initially empty double dot systems for our consideration.  This condition
dictates that the chemical potential of the drain electrodes should be
slightly lower than the ground state energy levels of the two QDs. 
Furthermore, since exchange coupling decreases as the level offset of the two
QDs increases \cite{BLD}, we will assume that the two QDs are identical with
the same gate voltage, so that the ground single electron energy levels of
the two dots are always aligned in order to maximize the exchange coupling. 
The gate voltages will be fixed to streamline the pulse sequence used to 
manipulate this double dot device.
To make the two-electron singlet state of the double dot accessible in
transport, the chemical potential of the source electrode has to be
higher than the singlet state energy of the double dot with two electrons
(but lower than the energy level of the two-electron triplet state, as we
discuss below).  In short, we require the source-drain bias voltage to enable
the system to satisfy the following inequalities: 
\begin{equation}
E(2e, {\rm triplet}) - E(1e) > \mu_{\rm source} > E(2e, {\rm singlet}) - E(1e)
\gtrsim E(1e) > \mu_{\rm drain} \,,
\end{equation}
%
%
where $E(2e)$ ($E(1e)$) is the lowest energy required to put two (one)
electrons in the double dot (defined relative to the bottom of the quantum dot
potential well).  Therefore, the source-drain bias voltage should be larger
than $E(2e, {\rm singlet})-E(1e)$ (which is generally larger than $E(1e)$
because the direct Coulomb repulsion caused increase is usually larger than
the exchange caused decrease in energy) if we align the bottom of the double
dot with the chemical potential of the drain electrodes, so that electrons
can tunnel into the two-electron singlet from the source electrode and then
empty into the drain electrodes.  Notice that lower bias voltages, though not
useful for the purpose of creating spin entangled electrons, may also lead to
physically interesting scenarios.  Under a small bias, the double dot can be
considered to be a large single dot.  When one excess electron occupies the
double dot, the two-dot charging energy would prevent another electron from
entering the double dot in the low-bias situation.  The system is essentially
in the Coulomb blockade regime of a single QD (with a relatively small
charging energy).  However, since there are two drain electrodes instead of
one, when the two paths recombine (for example, at a Y-junction), electron
interference effects such as Aharonov-Bohm effect can be expected as long as
the system size is smaller than the electron phase coherence length.  

The bias voltage needs to be small so that only one electron orbital level in
each quantum dot is open to accept electrons from the source and to emit
electrons to the drain.  If more quantum dot orbital levels are accessible to
the source electrode, electrons could tunnel into triplet and excited singlet 
states.  The occupation of these excited states would reduce the spin
filtering efficiency of the double dot.  For example, if a pair of electrons
are in a doubly-occupied singlet state (both electrons on the same QD),
they would have higher probability for entering the same drain electrode where
they would not represent a pair of distinguishable entangled electrons. 
These two electrons would also tend to have different energies so that their
control outside the double dot becomes problematic.  If the two electrons are
in a triplet state, they would either leave the quantum dots in a triplet
state, or undergo spin flips and possibly leave the double dot in a singlet
state, though this probability is only $33\%$ after one spin flip:
\begin{eqnarray}
& & |\uparrow\uparrow\rangle \rightarrow |\uparrow\downarrow\rangle
= (|T_0\rangle + |S\rangle)/\sqrt{2}, \nonumber \\
& & |\downarrow\downarrow\rangle \rightarrow |\uparrow\downarrow\rangle
= (|T_0\rangle + |S\rangle)/\sqrt{2}, \nonumber \\
& & |T_0\rangle \rightarrow (|\uparrow\uparrow\rangle
+ |\downarrow\downarrow\rangle)/\sqrt{2}\,. \nonumber
\end{eqnarray}
Since spin flip (or relaxation) time is quite long in QDs at low temperatures
(e.g. spin-flip tunneling time in the order of tens of
$\mu$s\cite{Fuji2,Hanson}), one cannot count on spin relaxation to bring a
pair of electrons in a triplet state into the ground singlet state within
their residence time inside the QDs (which we will show below to be in the
order of 100 ps).  Therefore, the bias voltage applied to the double quantum
dots has to be small: $V_{\rm bias} < E(2e, {\rm triplet})-E(1e)$, where
$V_{\rm bias} \equiv \mu_{\rm source} - \mu_{\rm drain}$.

The strength of dot-electrode tunneling should be sufficiently weak so that
second or higher order tunneling processes (cotunneling) are strongly
suppressed, because these higher order processes do not obey the spin
selection rule we established by properly setting the bias voltage (so that
only ground two-electron singlet state can pass through the double dot).
Indeed, we believe that cotunneling would make up of a part of the current
noise in the two streams of spin-entangled electrons, complicating the
observation of the entanglement induced cross-correlation and noise in the
current.

The source electrode is taken to be unpolarized (in an external magnetic field
this is not necessarily the case; for example, quantum Hall edge states have
been used to produce spin polarized injection into quantum
dots\cite{polarized}), so that any arbitrarily picked two electrons should
have a spin density matrix (neglecting the exchange interaction between the
two electrons and all spin-flip interactions) in which the two-spin
singlet state is equally probable as any of the three triplet states:
\begin{equation}
\rho = \frac{1}{4} (|S\rangle\langle S| + |T_0\rangle\langle T_0|
+ |T_{\uparrow}\rangle\langle T_{\uparrow}| 
+ |T_{\downarrow}\rangle\langle T_{\downarrow}|) \,.
\end{equation}
It is clear from this density matrix that the double QD acts here as a spin
filter that selectively allows the electron spin singlet component to tunnel
through.  In this sense the double QD system may be considered to be a
two-spin blockade device (in contrast to the regular spin blockade
devices,\cite{Ono,Huttel,Ciorga1} which have already been studied in the
literature).

\section{Problems with a fixed barrier double dot as a spin entanglement
filter}
\label{sect-problem}

When the conditions outlined in the previous section (which should be taken as
the minimal necessary conditions, by no means sufficient since, for example,
decoherence must be held to a minimum in the whole set-up so that
entanglement can be preserved) are all satisfied, the ground two-electron
singlet state becomes accessible to electrons tunneling through the double
QD.  However, a variety of tunneling events can occur through the double dot. 
It is thus crucial to analyze the relative weight of these tunneling events
in order to determine whether spin entangled states would be a dominant
component in the outgoing electron streams.

\begin{figure}
\centerline{
\epsfxsize=4.0in
\epsfbox{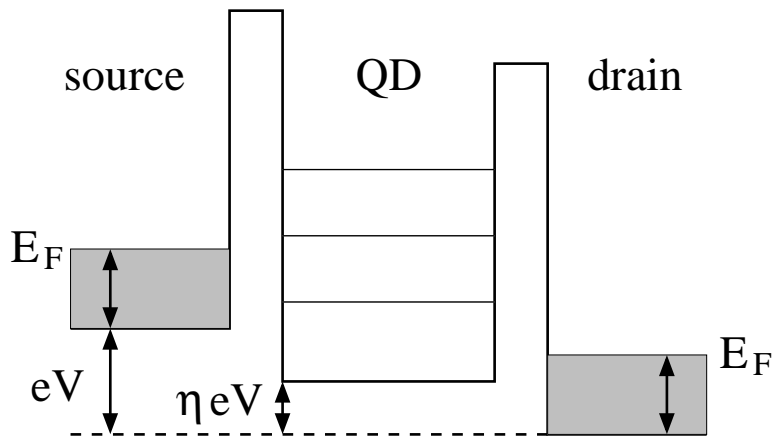}}
\vspace*{0.1in}
\protect\caption[]
{\sloppy{
Schematic diagram of a quantum dot in between a source and a drain electrode.
}}
\label{fig-S_QD_D}
\end{figure}

According to Ref.~\onlinecite{Beenakker}, the stationary current passing
through a QD (treating our coupled double dot as a single QD, as illustrated
in Fig.~\ref{fig-S_QD_D}) can be written as
\begin{equation}
I=-e \sum_p \sum_{\{n_i\}} \Gamma^{s}_p \, P(\{n_i\}) \,
\left\{ \delta_{n_p,0} f(E^{i,s}(N)-E_F) - \delta_{n_p,1}
\, [1-f(E^{f,s}(N)-E_F)] \right\} \,,
\end{equation}
where $V$ is the bias voltage applied across the QD, $p$ is the index of
orbital energy levels inside the QD, $\{n_i\}$ represents dot
level occupations, $\Gamma^{s(d)}_p$ are the tunneling matrix elements
between level $p$ and the source (drain) electrode, $P(\{n_i\})$ is the
stationary probability for configuration $\{n_i\}$ given by a set of detailed
balance equations, and $f$ is the Fermi-Dirac distribution.  $E^{i,s}(N)=E_p
+ U(N+1)- U(N) +\eta eV$ [$E^{f,s}(N)=E_p + U(N)- U(N-1) +\eta eV$] is the
energy of the source reservoir states from (to) which an electron tunnels
into (out of) state $p$ of a QD with $N$ electrons, where $E_p$ is the energy
of a single electron state $p$, $U(N)$ is the Coulomb interaction energy for
$N$ electrons in the QD, and $\eta eV$ is the voltage drop across the
QD-source-electrode potential barrier, as shown in Fig.~\ref{fig-S_QD_D}. 
In the simplified case we are considering here (neglecting occupation of any
configuration with $N>2$), the only relevant electron distribution
probabilities are $P(0)$, $P(1)$, and $P(2)$ (with zero, one, and two
electrons (singlet state) in the QD, and neglecting the two-electron triplet
states).  These three probabilities satisfy the detailed balance equations
for the individual configurations.\cite{Beenakker}  After some algebra, we
arrive at (for simplicity, assuming $T \rightarrow 0$) the occupation
probabilities
\begin{eqnarray}
P(0) & = & \frac{1}{1 + \Gamma^{s}_1/\Gamma^{d}_1 + \Gamma^{s}_1
\Gamma^{s}_2/\Gamma^{d}_1 \Gamma^{d}_2} \,, \nonumber \\
P(1) & = & \frac{1}{1 + \Gamma^{s}_2/\Gamma^{d}_2 + \Gamma^{d}_1
/\Gamma^{s}_1} \,, \nonumber \\
P(2) & = & \frac{1}{1 + \Gamma^{d}_2/\Gamma^{s}_2 + \Gamma^{d}_1
\Gamma^{d}_2/\Gamma^{s}_1 \Gamma^{s}_2} \,,
\label{eq:2e_probability}
\end{eqnarray}
and a current
\begin{equation}
I=e[\Gamma^d_1 P(1) + \Gamma^d_2 P(2)] \,.
\end{equation}
It is clear from Eq.~\ref{eq:2e_probability} that the probability of having
two electrons in this QD is very sensitive to the ratio
$r=\Gamma^{d}_2/\Gamma^{s}_2$.  If the tunneling rate to the drain electrode
$\Gamma^{d}_2$ is large, so that $r$ is large, $P(2)$ would be small: the QD
is then almost always empty.  Conversely, if the tunneling rate from the
source electrode $\Gamma^{s}_2$ is large, so that $r$ is small, $P(2)$ would
be large: the QD is almost always filled with two electrons.  For our purpose
of creating entanglement between electrons, which exists only in the
molecular two-electron state, we need to require $r \ll 1$ (equivalently, a
small $\Gamma^{d}_2$).  In this case, the current though the QD would be
dominated by $\Gamma^d_2 P(2)$, which corresponds to precisely one electron
tunneling out of the two-electron singlet state.

To use the double dot as a source of spin-entangled electrons, the pairs of
electrons have to leave the double dot with a certain degree of
synchronization.  However, as we have shown above, the electrical current
through the double dot is dominated by single electrons tunneling out of the
two-electron state.  The two-electron tunneling current, as a higher order
tunneling event, is proportional to $\Gamma^d_2 \Gamma^d_1 / U(2)$, where
$U(2)$ is the two-electron Coulomb repulsion energy and is of the same order
of magnitude as the single particle excitation energy and/or the external
bias potential ($eV$).  Since we require the level broadening $\hbar\Gamma$
to be small in order to achieve state-selective tunneling, the two-electron
current is much smaller than the one-electron current in our double dot
configuration, especially for small $\Gamma^d_2$ and $\Gamma^d_1$.

We thus face the problem of a conflicting dichotomy in the fixed barrier
double dot system in the sequential tunneling regime: it is difficult to
simultaneously have a highly occupied two-electron molecular state and a high
degree of synchronization for the electron pairs tunneling out of the double
dot.  Indeed, the output current is dominated by single electrons tunneling
out of the two-electron molecular state, which would render the double dot
useless as a spin entangler because the electrons coming out in transport
current are entangled with electrons that are localized in the QDs and are
therefore not accessible in transport measurements outside the double dot
system.  This problem needs to be addressed in the context of using the
double dot system as a spin entangler, and this is what we do in the next
Section.

\section{Synchronization of output from a double dot: A parallel 
Turnstile}
\label{sect-turnstile}

\begin{figure}
\centerline{
\epsfxsize=6.0in
\epsfbox{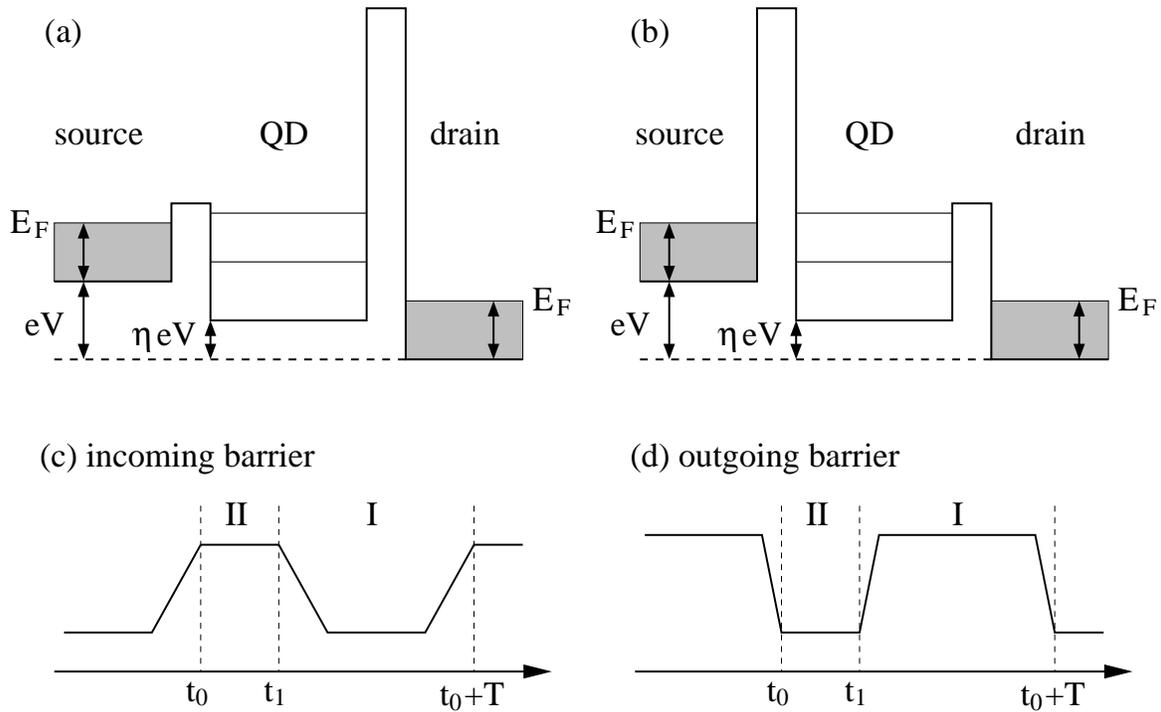}}
\vspace*{0.1in}
\protect\caption[]
{\sloppy{
The two stages of the double dot turnstile.
}}
\label{fig-turnstile}
\end{figure}

It is clear from the previous sections that the key problem facing fixed
barrier double quantum dot from the perspective of a spin entangler is the
difficulty in {\it controlled (regulated) extraction} of pairs of entangled
electrons.  One way to overcome the output synchronization problem is to
introduce time-dependent incoming and outgoing barriers for the double dot,
particularly the outgoing barrier.  This proposed introduction of
time-dependent barriers for the double QD is an analog of the single electron
turnstiles studied a decade ago.\cite{Kouwen1}  Within each cycle of the
turnstile, there are two stages of operation as illustrated in
Fig.~\ref{fig-turnstile}.  These two stages of operations would enable us to
avoid the conflicting dichotomy as each stage could optimize the individual
constraints discussed in the last section.  Specifically, during stage-I
(Fig.~\ref{fig-turnstile}a), the incoming barrier is low while the outgoing
barrier is high, so that two electrons would occupy the double dot and form a
spin singlet state.  During stage-II (Fig.~\ref{fig-turnstile}b), the
incoming barrier is high and the outgoing barrier is completely removed, so
that both electrons would rapidly ``empty'' into the two drain electrodes. 
The duration of the first stage is determined by the tunneling rate from the
source electrode into the QDs, while the duration of the second stage would
be determined by how fast the electrons diffuse out from the double dot into
the drain electrodes.

To ensure synchronization for the entangled electron pairs entering the drain
electrodes, the lowering of the outgoing barrier should be faster than the
interdot tunneling rate.  Otherwise, if one of the electrons tunnels out to
the drain during the barrier lowering period, the other electron may have
time to undergo charge oscillations between the two dots, therefore
significantly increasing the probability that it tunnels into the same drain
electrode as the other electron.  Such undesirable ``bunching'' events, where
two electrons enter the same drain electrode, would become part of the
current noise in the electrical current consisting of spin-entangled electron
pairs.  In realistic semiconductor QDs, this condition can be barely
satisfied with the best presently available microwave pulse generators.  For
example, if the exchange coupling between the double dot is 0.01 meV ($\sim
100$ mK, so that the experimental temperature has to be much lower than 100
mK in order to avoid tunneling through the two-electron triplet states),
which is a reasonable number,\cite{HD1,HD2} and the on-site Coulomb repulsion
energy is about 1 meV (again a typical number), the inter-dot tunnel coupling
would have to be in the order of 0.1 meV, corresponding to a tunneling time
of about 40 ps.  The pulse ramp up time of the present-day pulse generators
is about 30 to 40 ps, comparable to the tunneling time.  To improve the
quality of this double dot electron spin entangler, sharper pulses are
desirable.  

Current up to 1 nA is needed to evaluate current noise and cross correlation
with the state-of-the-art technology.  For a 1 nA current, the electrons would
stay in the QD on the average for about 160 ps.  In the case of a turnstile,
the corresponding repetition frequency is about 6 GHz.  The inter-dot
coupling can then be about 20 GHz (tunneling time of 50 ps) to ensure that
electrons would form molecular states inside the double dot.  The barrier
varying pulse width should be about 20 ps (shorter than the inter-dot
tunneling time), which is at the limit of the present technology.\cite{Fuji1} 

%
%
An important question regarding the operation of a double dot turnstile
structure as envisioned here is whether the fast gate voltage pulses would
cause electron orbital or spin states to get excited.  To avoid such
unintended excitations, the gate voltages controlling the incoming and
outgoing barriers should be varied synchronously so that the overall effect
on the electrons inside the double quantum dot remains small.  Adiabatic
condition can then be satisfied and the electron states would remain
unchanged.  Quantitatively, adiabatic condition requires that $|dV/dt|/V <<
J/\hbar$.  The fastest rising voltage pulses have a rise time of about 30 ps,
corresponding to $|dV/dt|/V \sim 30$ GHz.  For $J=0.1$ meV, $J/\hbar \sim
100$ GHz, barely satisfying the adiabatic condition above.  Increasing $J$
would make the adiabatic condition easier to satisfy.  Furthermore, if
spin-flip interaction (such as spin-orbit coupling) is sufficiently weak in
the double dot system, the only possible excitation caused by gate voltage
variation would be changes in the electron orbital states.  In this case the
exchange $J$ in the inequality above should be replace by the single particle
excitation energy $E$ that is generally at least one order of magnitude
larger than $J$.  Therefore, the fast voltage pulses required to operate the
double dot turnstile should in general not cause serious problem in terms of
exciting electrons into unwanted states.

\section{Electron entanglement detection using a beam splitter:
Consequences of wavepacket states}
\label{sect-wavepacket}
%
%

Assuming spin-entangled electron pairs can be generated synchronously with
sufficient consistency from a double dot entangler (following considerations
discussed in the earlier sections), the important question arises as how
to measure and quantify the degree of entanglement in the electron pairs.
One approach that can take advantage of ensemble-averaging is to use a
beam splitter to create two-electron interference and then measure the noise
and correlation spectra of electrical currents.\cite{BLS}  Here we would
like to explore whether this approach can work effectively with electron
pairs coming out of the double dot entangler in the turnstile configuration
discussed in the last section.

\begin{figure}
\centerline{
\epsfxsize=4.0in
\epsfbox{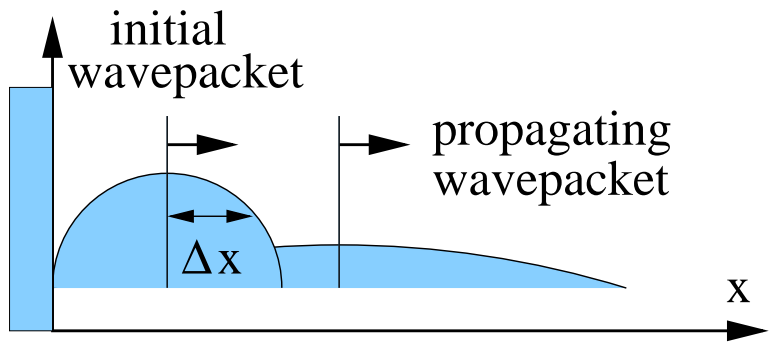}}
\vspace*{0.1in}
\protect\caption[]
{\sloppy{
Free expansion of the electron wavepackets during stage-II of the trunstile
operation.  The shaded area to the left of the wavepackets represents the
barrier that separate the quantum dots from the source electrode.  The barrier
between the QDs and the drain electrodes is removed so that the QDs are now
completely open, which results in the free expansion of the electron
wavepackets. 
}}
\label{fig-wavepacket}
\end{figure}

In Ref.~\onlinecite{BLS}, the electrons injected into the beam splitter are
assumed to be in plane wave states or purely momentum states.  If a double QD
works as an electron spin entangler, the electrons coming out of the double
dot are in expanding or dispersing wavepacket states.  The use of partly
localized wavepacket states instead of extended plane wave states also
highlight the contrast between the quantum mechanical dual wave-particle
nature of the electrons and their classical particle-like image.  As shown in
Fig.~\ref{fig-turnstile}, in stage-II of each turnstile cycle the outgoing
barrier for the double QD is lowered to facilitate the dot-emptying process.  
In the simple limit that the outgoing barrier is lowered all the way so that
the electron wavefunction will undergo free expansion {\it without tunneling},
as illustrated schematically in Fig.~\ref{fig-wavepacket},
the peak of the wavepacket would propagate along the drain electrode
(which is assumed to be one-dimensional for simplicity) with a speed of
$\pi\hbar/ma \propto \sqrt{E_{\rm conf}}$ with $a$ being the confinement
length and $E_{\rm conf}$ the confinement energy, and the width of the
wavepacket would expand with a speed $\propto \sqrt{1+(E_{\rm conf}
t/\hbar)^2}$.\cite{Schiff}  For example, for a confinement length of 100 nm,
which corresponds to a confinement energy $\sim 0.5$ meV, the wavepacket
propagates with a speed of 1 $\mu$m per 10 ps, within which period the
wavepacket becomes 10 times wider.  Notice that the electron dynamics here is
quite different from the case of tunneling.  Energy spread of an electron
wavepacket would be determined by the tunneling rate $\Delta \epsilon \sim
\hbar \Gamma$ if the electron tunnels through a barrier out of the quantum
dot, no matter whether the system is in the turnstile or fixed barrier
configuration.  In our turnstile proposal here, the outgoing barrier is
lowered all the way to zero during stage-II of the turnstile operation
(Fig.~\ref{fig-turnstile}), so that the confined electron can expand out of
the initial state freely into the lead without facing any barrier
(Fig.~\ref{fig-wavepacket}).  The energy of such an electron initially is
$E=\langle H \rangle=E_{\rm conf}$.  The energy spread is determined from the
initial spatial spread $\Delta x$, which is also related to the initial
confinement energy $E_{\rm conf}$ as we discussed above.

We now focus on the observability of the difference between electron singlet
and triplet states assuming the electrons coming into the beam splitter (which
is treated as a four-terminal device, as illustrated in Fig.~\ref{fig-Sch})
are in wavepacket states.  The current noise in each of terminals 3 and 4 and
the cross correlation between currents in terminals 3 and 4 can be written
as:\cite{BLS,Buttiker}
\begin{equation}
S_{\alpha \beta} = \lim_{T \rightarrow \infty} \frac{1}{2\pi T} \int_0^T
dt \ e^{i \omega t} \langle \psi |\delta I_{\alpha}(t) \delta
I_{\beta}(0) | \psi \rangle \,,
\end{equation}
where $\alpha \,,\beta = 3\,,4$, and the current can be expressed in terms of
the creation and annihilation operators at the terminals:\cite{Buttiker}
\begin{equation}
I_{\alpha \sigma} (t) = \frac{e}{h} \sum_{E E'} \left[
a_{\alpha \sigma}^{\dagger}(E) a_{\alpha \sigma}(E') 
- b_{\alpha \sigma}^{\dagger}(E) b_{\alpha \sigma}(E') \right]
e^{i(E-E^{\dagger})t/\hbar} \,.
\label{eq:current}
\end{equation}
According to Ref.~\onlinecite{BLS}, if the electron pairs are initially in
plane wave states with energies $\epsilon_1$ and $\epsilon_2$: 
\begin{equation}
|\pm\rangle = \frac{1}{\sqrt{2}} \left[ a_{2 \downarrow}^{\dagger}
(\epsilon_2) a_{1 \uparrow}^{\dagger}(\epsilon_1) \pm a_{2
\uparrow}^{\dagger}(\epsilon_2) a_{1 \downarrow}^{\dagger}(\epsilon_1)
\right] |0\rangle \,,
\end{equation}
the outgoing current noise and cross correlation from the beam splitter in
leads 3 and 4 satisfies\cite{BLS}
\begin{equation}
S_{\alpha \beta} \propto T(1-T)(1 \mp \delta_{\epsilon_1,\epsilon_2}) \,,
\label{eq:noise}
\end{equation}
depending on whether the initial spin state is one of the triplet states or
the singlet state.  The physical picture of this difference is conceptually
simple to understand.  The two-electron singlet entangled state is
antisymmetric in spin space and therefore symmetric in spatial coordinates
between the two electrons (in order to maintain the overall antisymmetric
nature of the two-electron fermionic state).  This symmetric real space nature
of the orbital two-electron wavefunction of the singlet pair then introduces
the differences in the cross-correlations and the current noise spectra as
the entangled electrons scattering off the beam splitter tend to ``bunch"
together.  The observation of the noise and cross-correlation in the output
currents of the beam splitter can thus be construed as a direct (transport)
evidence for spin entanglement created back in the double quantum dot.  Here
we comment that such an interference type experiment is a solid state analog
of the famous Hanbury Brown and Twiss experiment for photons\cite{HBT} and a
natural extension of similar solid state
experiments\cite{Liu,Oliver,Maitre,Ji} with entangled electrons.

It is clear that the $\delta$-function in Eq.~\ref{eq:noise} is a direct
consequence of the choice of the initial plane wave states for the electrons.
A more realistic choice in our case, where electrons expand out of the quantum
dots during stage-II of the turnstile cycles (see Fig.~\ref{fig-turnstile}),
would be the wavepacket states.  If the orbital part of the initial wavepacket
takes a simple Gaussian form:
\begin{equation}
\Psi(x,t) = \frac{1}{(2\pi)^{1/4} \sqrt{\Delta x}} \ e^{-\frac{(x - \langle x
\rangle)^2}{4 (\Delta x)^2} + i\frac{\langle p \rangle}{\hbar}x}  \,,
\end{equation}
where $\Delta x$ is the wavepacket width, and the average momentum $\langle p
\rangle$ of the wavepacket is determined by the electron energy $E$ inside
the QD and the Fermi energy in the drain electrode.  The initial electronic
state entering the beam splitter would now take on a more complicated form:
\begin{eqnarray}
|\pm\rangle & = & \int_{0}^{\infty} dk_1 \int_{0}^{\infty} dk_2 \ \phi_1 (k_1)
\phi_2 (k_2) \times \frac{1}{\sqrt{2}} \left( a_{2 \downarrow}^{\dagger}
(k_2) a_{1 \uparrow}^{\dagger}(k_1) \pm a_{2 \uparrow}^{\dagger}(k_2) a_{1
\downarrow}^{\dagger}(k_1) \right) |0\rangle \nonumber \\
& = & \int_{0}^{\infty} d\epsilon_1 \int_{0}^{\infty} d\epsilon_2 \ 
\psi_1(\epsilon_1) \psi_2(\epsilon_2) \times \frac{1}{\sqrt{2}} \left( a_{2
\downarrow}^{\dagger} (\epsilon_2) a_{1 \uparrow}^{\dagger}(\epsilon_1) \pm
a_{2 \uparrow}^{\dagger}(\epsilon_2) a_{1 \downarrow}^{\dagger}(\epsilon_1)
\right) |0\rangle \,,
\end{eqnarray}
where
\begin{equation}
\phi_i (k) = \frac{2}{(2\pi)^{1/4}} \sqrt{\frac{\Delta x}{L}} \ e^{-(k-k_i)^2
(\Delta x)^2} e^{i\theta_i}\,.
\end{equation}
Here $\theta_i = E_k t_i/\hbar$ is an initial phase determined by when the
wavepacket enters the lead.  Notice that there is no time-dependence in this
wavepacket wavefunction because we are working in the Heisenberg picture (see
Eq.~(\ref{eq:current})).  Physically, the wavepacket state is made up of a
continuous spectrum of plane wave states, and represents a spatially
localized electron.  
%
%
Going through similar type of algebra as in Ref.~\onlinecite{BLS} using
wavepacket states rather than plane wave states, we arrive at the expression
for the current noise and cross correlation:
\begin{equation}
S_{\alpha \beta} \propto T(1-T) \left( 1 \mp \left|\int dk  \
\phi_1^*(k) \phi_2(k) \right|^2 \right) \,.
\end{equation}
Thus, in the case of a simple Gaussian wavepacket, the noise/cross-correlation
function is given by (with $\alpha$ and $\beta$ being 3 or 4), 
\begin{equation}
S_{\alpha \beta} \propto T(1-T) \left[ 1 \mp \frac{1}{1+d^2/4} \ 
e^{-\frac{1}{1+d^2/4}(k_1 - k_2)^2 (\Delta x)^2} \ 
e^{-\frac{d^2}{2+d^2/2}(k_1^2 + k_2^2) (\Delta x)^2}\right] \,,
\label{eq:wp_noise}
\end{equation}
where $d = \hbar (t_1 - t_2)/2m(\Delta x)^2$ represents the difference
between the initial phases of the two wavepackets (when they each entered the
leads), and $k_1 \sim \langle p \rangle_1 \propto \sqrt{\epsilon_1}$ and $k_2
\sim \langle p \rangle_2 \propto \sqrt{\epsilon_2}$.  The strength of the
signature of electron spin singlet/triplet states is thus determined by the
spectral overlap of the two electron wavepackets coming into the beam
splitter relative to the spectral width of each individual wavepacket, and
when they each enter the leads (therefore the beam splitter). 
%
%
In the double dot turnstile configuration, $\epsilon_i$'s are determined by
the energy levels inside the dots (assuming the two drain electrodes have the
same Fermi energy).  The energy levels of the two quantum dots need to be
aligned not only to maximize the interdot exchange coupling $J$, but also
keep $k_1 - k_2$ to be relatively small.  $\Delta x$ takes on the value of
confinement length in the case of a turnstile when the outgoing barrier is
completely removed during stage-II in Fig.~(\ref{fig-turnstile}).   It can be
estimated that for reasonable initial energy differences the
$[(k_1-k_2)\Delta x]^2$ factor in Eq.~(\ref{eq:wp_noise}) should be smaller
than one so that the first exponential factor in Eq.~(\ref{eq:wp_noise})
should not suppress the interference signal by much.  If the electrons enter
the beam splitter (thus the drain electrodes) at the same time, so that $d
\propto t_1-t_2 = 0$, the expression in Eq.~\ref{eq:wp_noise} can be
simplified:
\begin{equation}
S_{\alpha \beta} \propto T(1-T) \left[ 1 \mp \ 
e^{-(k_1 - k_2)^2 (\Delta x)^2} \right] \,.
\label{eq:wp_noise1}
\end{equation}
In this case the contrast between entangled (singlet) and non-entangled
electron pairs should be mostly preserved, so that plain wave description of
the electrons\cite{BLS} should work reasonably well.  If the electrons do not
enter the beam splitter simultaneously due to reasons such as differences in
the two outgoing barriers or difference in the lengths of the electrodes
between the QDs and the beam splitter, the effect of the initial phase
difference is non-negligible.  For example, for $\Delta x \sim 10$ nm, $d
\sim 10^{13} (t_1 - t_2)$.  If $t_1-t_2$ is in the order of 10 picosecond,
which could be the case if the outgoing barrier of the double dot turnstile
is modulated on the time scale of 30 picosecond, the factor $d$ satisfies $d
\gg 1$.  This re-enforces the fact that the first exponential factor in
Eq.~(\ref{eq:wp_noise}) is not very important in the present case (because of
the $1/(1+d^2/4)$ factor in the exponent).  The exponent of the second
exponential factor can be approximated by $2(k_1^2+k_2^2)\Delta x^2$, which
is in the order of 1 for $\Delta x \sim 10$ nm and the average electron
energy $\sim 1$ meV.  If electrons have larger kinetic energy, the difference
between singlet and triplet states would be significantly suppressed in the
signal from the beam splitter. 

When the electrons come from a fixed barrier double dot, the spectral width
$\Delta\epsilon$ for each wavepacket is determined by the decay rate through
the outgoing potential barrier, which is much smaller than the confinement
energy ($\hbar \Gamma \ll E_{\rm conf}$ so that bound states in the QDs are
well defined).  Now $\Delta x$ would generally be very large (for a dot
lifetime of 1 ns, $\Delta x$ could be larger than 10 $\mu$m), so that
$d = \hbar (t_1 - t_2)/2m(\Delta x)^2$ is small even for $t_1 - t_2$ in
the order of ns (here we neglect effects from electrons not entangled, which
would themselves cause deterioration in the signal).  Now the first
exponential factor in Eq.~(\ref{eq:wp_noise}) would dominate, so that even a
small energy difference of 0.01 meV can cause significant signal suppression.  
This situation is a direct extension of treating electrons as plane waves, for
which the spectral width $\Delta \epsilon$ goes to zero.  The result here
shows again the advantage of the double QD turnstile configuration
as compared to the fixed barrier QDs from the perspective of the signal to
noise ratio in this particular beam splitter detection scheme.

If the beam splitter is too far downstream from the double dot
entangler, the wavepackets would lose their phase coherence through
scattering, and the two-electron spin state deteriorates through spin
decoherence and relaxation, so that the strength of the
signature of singlet/triplet states would reduce accordingly.\cite{BL}

\section{Discussion}

With a double dot turnstile, only the singlet spin-entangled state
$|\uparrow\downarrow\rangle - \downarrow\uparrow\rangle$ can be filtered from
an unpolarized reservoir.  Although an external magnetic field can make
the triplet state the ground state of a double dot,\cite{BLD,HD1,HD2} the
lowest energy state of the triplet would be a polarized unentangled state
(e.g. $|\uparrow\uparrow\rangle$).  To generate the other three Bell states
(maximally entangled two-spin states: $|\uparrow\downarrow\rangle +
\downarrow\uparrow\rangle$ and $|\uparrow\uparrow\rangle \pm
|\downarrow\downarrow\rangle$), the double dot turnstile must be combined
with single spin operations such as phase shift and spin flip.  

One potential problem with the turnstile configuration is the unintended
inter-gate cross talk because of the simultaneous presence of closely-packed
gates (needed to produce small quantum dots so that the exchange coupling
between the two dots is sizable) and the relatively high operational
frequency for the turnstile (6 GHz).  From a control point of view, the
residence time for the electrons inside the quantum dots should be as long as
possible ($> 1$ ns) in order to lower the operational frequency of the
turnstile.  However, the corresponding smaller current would lead to a
suppression of the signal-to-noise ratio in the final measurement,
particularly if the measurement device is a beam splitter, since
current noise and correlation measurement require a relatively large current
(in the order of nA).  Future development in the current measurement
techniques would certainly help the experimental demonstration of the double
dot turnstile entangler through the beam splitter detection scheme.

%
%
Another question regarding the turnstile scheme is whether rapidly varying
potential barriers would significantly increase the energy discrepancy
between the two electrons as they leave the quantum dot for the leads.  A
simple estimate can help clarify this issue.  The fastest pulse rise/fall
time available now is about 30 ps.  Using the uncertainty principle, the
corresponding energy uncertainty is smaller than 0.1 meV.  Recall that $\Delta
x$ is in the orders of $10 \sim 100$ nm, the exponent $(\Delta k \Delta x)^2$
in Eq.~\ref{eq:wp_noise} can be estimated to be in the order of $10^{-2} \sim
10^{-4}$ if $E_{\rm conf}$ is $\sim 1$ meV.  Similarly, the added fluctuation
in energy would not cause significant increase in $(k_1^2 + k_2^2)\Delta x^2$
either because the wavepacket energies $\epsilon_i=\hbar^2 k_i^2/2m$ are
already in the order of meV to begin with.  Therefore, the energy
fluctuations caused by the fast pulses operating the turnstile should not
lead to any significant suppression of the current noise and correlation
signal after the beam splitter.

In this paper we studied how the efficiency of a beam splitter is reduced
because of the wavepacket nature of the conduction electrons.  There are also
other physical processes that lead to decreased sensitivity for a beam
splitter.\cite{BL,Egues}  For example, it has been shown that the presence of
spin-orbit coupling in the vicinity of a beam splitter, which is particularly
true for semiconductor nanostructures based on two-dimensional electron gas
(2DEG), can hurt the detection efficiency of a beam splitter.\cite{Egues} 
Indeed, the asymmetric nature of the 2DEG and the associated spin-orbit
coupling pose a potential problem to the general study of spin-based quantum
computing near interfaces and need to be further studied.

The most important issue in the observability of electronic spin entanglement
in this context is perhaps the problem of decoherence, which is usually
severe for electrons in solid state systems.  The question is the extent to
which spin entanglement, even if it is successfully produced in the double
dot spin entangler, will survive through the outgoing leads to be detected in
the current noise correlation measurements.  One would want to have the
temperature as low as possible and the transport should be ballistic through
the leads in order to minimize phase breaking scattering.  Ideally one
desires current leads which are devoid of electrons themselves (e.g. carbon
nanotubes\cite{Bena}), but are capable of conducting current so as to
eliminate electron-electron interaction effects as much as possible.  In
reality, the leads are likely to be metallic, and one must make them
reasonably short in order not to lose entanglement as the electrons
transverse through the leads.\cite{BLS}  Whether this can be achieved or not
experimentally is unknown at this stage.

\section{Conclusion}

In this paper we study the conditions necessary for a double quantum dot
system to work as an efficient electron spin entangler.  We have analyzed in
detail the required relative strength of the tunneling matrix elements, and
the desirable bias and gate voltages.  The difficulties with fixed barrier
double dots, especially with extracting electron pairs, are carefully
clarified, thus establishing the necessity of time-dependent input/output
barriers which would enable optimization of the various tunneling rates
needed for producing entanglement and for regularly exporting pairs of
spin-entangled electrons.  We discuss the conditions on the
turnstile configuration for the double dot to be an efficient generator of
spin-entangled electron pairs.  We also analyze the consequences of
wavepacket states for the electrons as they leave the double dot.  We 
show that a double quantum dot turnstile is, in principle, an efficient
electron spin entangler or entanglement filter because of the exchange
coupling between the dots and the controllable electron output that is
possible in such a device.  Whether electronic entanglement in a double
quantum dot system can be experimentally detected in a transport
noise/correlation measurement would depend on many conditions, including the
coherence in the output leads, but our work shows that, as a matter of
principle, such transport measurements are feasible.

We thank LPS, ARDA, and ARO for partial financial support.  We also thank
useful discussions with Y.Z. Chen and Richard Webb.


\begin{thebibliography}{99}
\bibitem{HF_comment} The mean field Hartree-Fock Slater determinant states of
electrons in solids (or atoms or molecules) cannot be factored into a product
of single particle states because they incorporate Pauli principle through
the appropriate antisymmetrization (thus reflect the fermionic
characteristics of electrons).  However, such correlation is not nonlocal. 
When the particles are separated spatially, the antisymmetrization does not
carry physical meaning anymore.  Therefore, single Slater determinant states
should not be considered entangled.\cite{Schliemann}
\bibitem{weak_int} If two entangled objects weakly interacts with each other,
but rapid measurement is available, Bell-type comparison of measurement
results may still be possible with high fidelity.\cite{Rowe}  However, the
measurement techniques (for spins, for example) are generally quite slow in
solids so that such scenario is generally difficult to realize.
\bibitem{spatial_sep} Observation of entanglement requires that the
interaction that creates the entangled state be turned off when the nonlocal
correlations are measured between the entangled pairs---there is nothing
unusual about finding correlations between objects that are strongly
interacting with each other.  For this very same reason interaction or
coupling induced mode repulsion or resonant level splitting, which is fairly
common in solid state spectroscopic measurements, is not direct evidence for
entanglement---such resonant level repulsion is only evidence for coupling
between two objects.  (In fact, such level repulsion by itself is not even
evidence for a quantum mechanical coupling between the two systems or energy
levels since two coupled classical oscillators give rise to qualitatively
identical mode repulsion or level splitting behavior.)
\end{thebibliography}
\end{document}